\documentclass[twocolumn,preprintnumbers,superscriptaddress,nofootinbib]{revtex4}
\usepackage{graphicx}
\usepackage{amssymb}
\usepackage{color}
\usepackage{enumerate}
\usepackage{bm}
\def\beq{\begin{equation}}
\def\eeq{\end{equation}}
\def\beqn{\begin{eqnarray}}
\def\eeqn{\end{eqnarray}}

\renewcommand{\texttt}{{}}
\newcommand{\be}{\begin{eqnarray}}
\newcommand{\ee}{\end{eqnarray}}
\newcommand{\bee}{\begin{equation}}
\newcommand{\eee}{\end{equation}}

\begin{document}

\title{Non-local massive gravity} 

\author{Leonardo Modesto}
\email{lmodesto@fudan.edu.cn}
\affiliation{Department of Physics \& Center for Field Theory and Particle Physics, \\
Fudan University, 200433 Shanghai, China}
\author{Shinji Tsujikawa}
\email{shinji@rs.kagu.tus.ac.jp}
\affiliation{Department of Physics, Faculty of Science, Tokyo University of Science, 
1-3, Kagurazaka, Shinjuku-ku, Tokyo 162-8601, Japan}

\date{\small\today}

\begin{abstract} \noindent

We present a general covariant action for massive gravity merging 
together a class of ``non-polynomial'' and super-renormalizable or finite 
theories of gravity with the non-local theory of 
gravity recently proposed by Jaccard, Maggiore and 
Mitsou (Phys.\ Rev.\ D {\bf 88} (2013) 044033). 
Our diffeomorphism invariant action gives rise to the  
equations of motion appearing in non-local massive massive gravity 
plus quadratic curvature terms.
Not only the massive graviton propagator reduces 
smoothly to the massless one without a vDVZ discontinuity, 
but also our finite theory of gravity is unitary at tree level 
around the Minkowski background.
We also show that, as long as the graviton mass $m$ is much smaller
the today's Hubble parameter $H_0$, a late-time cosmic acceleration 
can be realized without a dark energy component
due to the growth of a scalar degree of freedom.
In the presence of the cosmological constant $\Lambda$, 
the dominance of the non-local mass term leads to
a kind of ``degravitation''  for $\Lambda$ at the late 
cosmological epoch.

\end{abstract}

\maketitle

\section{Introduction} 

The construction of a consistent theory of massive gravity has a 
long history, starting from the first attempts of Fierz and Pauli \cite{Fierz} 
in 1939. The Fierz-Pauli theory, which is a simple extension of General 
Relativity (GR) with a linear graviton mass term, is plagued by 
a problem of the so-called van Dam-Veltman-Zakharov (vDVZ)
discontinuity  \cite{DVZ}. 
This means that the linearized GR is not 
recovered in the limit that the graviton mass is sent to zero.

The problem of the vDVZ discontinuity can be alleviated in the 
non-linear version of the Fierz-Pauli theory \cite{Vain}.
The non-linear interactions lead to a well behaved 
continuous expansion of solutions 
within the so-called Vainshtein radius.
However, the nonlinearities that cure the vDVZ discontinuity 
problem give rise to the so-called Boulware-Deser (BD) 
ghost \cite{BDghost} with a vacuum instability.

A massive gravity theory free from the BD ghost was constructed by de Rham, 
Gabadadze and Tolley (dRGT) \cite{dRGT}
as an extension of the Galileon gravity \cite{Galileon}.
On the homogenous and isotropic background, however,
the self-accelerating solutions in the dRGT theory exhibit 
instabilities of scalar and vector perturbations \cite{ins}. 
The analysis based on non-linear cosmological perturbations shows that 
there is at least one ghost mode (among the five degrees of 
freedom) in the gravity sector \cite{Antonio}.
Moreover it was shown in Ref.~\cite{aca} that the constraint 
eliminating the BD ghost gives rise to an acausality problem.
These problems can be alleviated by extending the original dRGT 
theory to include other degrees of freedom \cite{quasi1,quasi2,quasi3} 
(like quasidilatons) or by breaking the homogeneity \cite{homoge} 
or isotropy \cite{Gum,Antonio2} of the cosmological background.

Recently, Jaccard {\it et al.} \cite{Maggiore} constructed 
a non-local theory of massive gravity by using a quadratic action of 
perturbations expanded around the Minkowski background.
This action was originally introduced in Refs.~\cite{Arkani,Dvali} 
in the context of the degravitation idea of the cosmological constant.
The resulting covariant non-linear theory of massive gravity
not only frees from the vDVZ discontinuity but respects causality. 
Moreover, unlike the dRGT theory, it is not required to introduce 
an external reference metric.

Jaccard {\it et al.}  \cite{Maggiore}  showed that, on the Minkowski 
background, there exists a scalar ghost in addition to the five 
degrees of freedom of a massive graviton, by decomposing 
a saturated propagator into spin-2, spin-1, and spin-0 
components. For the graviton mass $m$ of the order of the 
today's Hubble parameter $H_0$, the vacuum decay rate
induced by the ghost was found to be very tiny even over 
cosmological time scales. 
The possibility of the degravitation of a vacuum energy was 
also suggested by introducing another
mass scale $\mu$ much smaller than $m$.

In this paper we propose a general covariant action principle 
which provides the equations of motion for the non-local massive 
gravity \cite{Maggiore} with quadratic curvature terms.
The action turns out to be a bridge between a class of 
super-renormalizable or finite theories of quantum
gravity \cite{Krasnikov, Tomboulis, M1, BM, M2, M3, M4} 
and a diffeomorphism invariant theory for a massive graviton.

The theory previously studied in Refs.~\cite{Krasnikov, Tomboulis, M1, BM, M2, M3, M4} 
has an aim to provide a completion of the Einstein gravity
through the introduction of a non-polynomial or semi-polynomial entire 
function (form factor) without any pole in the action. 
In contrast, the non-local massive gravity studied in this paper shows a pole 
in the classical action making it fully non-local. 
However, the Lagrangian for massive gravity can be selected out 
from the theories previously proposed \cite{Krasnikov, Tomboulis, M1, BM, M2, M3, M4}
once the form factor has a particular infrared behavior. 
The non-local theory resulting from the covariant Lagrangian is found to 
be unitary at tree level on the Minkowski background.
Moreover, the theory respects causality and smoothly reduces 
to the massless one without the vDVZ discontinuity. 

We will also study the cosmology of non-local massive gravity
on the flat Friedmann-Lema\^{i}tre-Robertson-Walker (FLRW) background
in the presence of radiation and non-relativistic 
matter\footnote{Note that cosmological consequences 
of non-local theory given by the Lagrangian $Rf(\Box^{-1}R)$ 
have been studied in 
Refs.~\cite{Deser,Jhingan,Koivisto,Woodard,Zhang,Elizalde,Park}. 
In this case the function $f(\Box^{-1}R)$ can be chosen only 
phenomenologically from the demand to realize the late-time 
cosmic acceleration and so on.}.
Neglecting the contribution of quadratic curvature terms 
irrelevant to the cosmological dynamics much below the Planck scale, 
the dynamical equations of motion reduce to those derived in 
Ref.~\cite{Maggiore}. We show that, as long as 
the graviton mass $m$ is much smaller than $H_0$, 
the today's cosmic acceleration can be realized 
without a dark energy component due to the growth of 
a scalar degree of freedom. 

Our paper is organized as follows.
In Sec.~\ref{theorysec} we show a non-local covariant Lagrangian
which gives rise to the same equation of motion as that in non-local massive 
gravity with quadratic curvature terms.
We also evaluate the propagator of the theory to study the tree-level unitarity.
In Sec.~\ref{cosmosec} we study the cosmological implications
of non-local massive gravity in detail to provide a minimal 
explanation to dark energy in terms of the graviton mass. 
We also discuss the degravitation of the cosmological constant
induced by the non-local mass term.
Conclusions and discussions are given in Sec.~\ref{consec}.

Throughout our paper we use the metric signature 
$\eta_{\mu \nu}={\rm diag}(+1,-1,-1,-1)$. 
The notations of the Riemann tensor, the Ricci tensor,
and Ricci scalar are 
$R^{\mu}\!_{\nu \rho \sigma} = - \partial_{\sigma} \Gamma^{\mu}_{\nu \rho} + \dots$, 
$R_{\mu \nu } = R^{\sigma}\!_{\mu \nu \sigma}$ and 
$R = g^{\mu \nu} R_{\mu \nu}$, respectively.

\section{Super-renormalizable non-local Gravity} \label{theorysec}

Let us start with the following general class of non-local actions 
in $D$ dimension \cite{Krasnikov, Tomboulis, M1, BM, M2, M3, M4}, 
\be 
&& \hspace{-0.5cm} 
S \! = \! \int d^D x \sqrt{|g|} \Big[2 \kappa^{-2} R + \bar{\lambda}
+ \, \overbrace{ O(R^3) \dots\dots\dots 
+ R^{\mathrm{N}+2}}^{\mbox{Finite number of terms}} 
\nonumber \\
&& 
+ \sum_{n=0}^{\mathrm{N}} \Big( 
a_n \, R \, (-\Box_{M})^n \, R  + 
b_n \, R_{\mu \nu} \, (-\Box_{M})^n \, R^{\mu \nu} \Big) 
\nonumber \\
&& + R  \, h_0( - \Box_{M}) \, R +
R_{\mu\nu} \, h_2( - \Box_{M}) \, R^{\mu \nu}
\Big] \,, 
\label{action}
\ee 
where $\kappa=\sqrt{32 \pi G}$ ($G$ is gravitational constant), 
$|g|$ is the determinant of a metric tensor $g_{\mu \nu}$, 
$\Box$ is the d'Alembertian operator with $\Box_M=\Box/M^2$, 
and $M$ is an ultraviolet mass scale. 
The first two lines of the action consist of a finite number of 
operators multiplied by coupling constants 
subject to renormalization at quantum level.
The functions $h_2(z)$ and $h_0(z)$, where $z \equiv - \Box_M$, 
are not renormalized and defined as follows
\be
&& \hspace{-0.5cm} 
h_2(z) = \frac{ V(z)^{-1} -1 - \frac{\kappa^2 M^2}{2} \, z 
\sum_{n=0}^{\mathrm{N}} \tilde{b}_n \, z^n}
{\frac{\kappa^2 M^2}{2}\, z} 
\, , \,\,\,\,\,  \nonumber \\
&& \hspace{-0.5cm} 
h_0(z) = - \frac{V(z)^{-1} -1 + \kappa^2 M^2 \, z
\sum_{n=0}^{\mathrm{N}} \tilde{a}_n \, z^n}{
\kappa^2 M^2 \, z} \,, 
\label{hzD}
\ee
for general parameters $\tilde{a}_n$ and $\tilde{b}_n$, while 
\be
&&\hspace{0cm}
V(z)^{-1} := \frac{\square + m^2}{\square} \, 
e^{{\mathrm  H}(z)}\,, \label{Vdef0} \\
&& e^{{\mathrm H}(z)} = \left| p_{\gamma + \mathrm{N} + 1}(z) \right|  
\, e^{\frac{1}{2} \left[ \Gamma \left(0, p_{\gamma + \mathrm{N} + 1}^2(z) 
\right)+\gamma_E  \right] }\,.
\label{Vdef}
\ee

The form factor $V(z)^{-1}$ in Eq.~(\ref{Vdef0}) is made of two parts: 
(i) a non-local operator $(\Box+m^2)/\Box$ 
which goes to the identity in the ultraviolet regime, and 
(ii) an entire function $e^{{\mathrm  H}(z)}$ without zeros 
in all complex planes. 
Here, $m$ is a mass scale associated with the graviton mass 
that we will discuss later when we calculate 
the two-point correlation function.
${\mathrm H}(z)$ is an entire function 
of the operator $z=-\Box_M$, and 
$p_{\gamma + \mathrm{N} + 1}(z)$ is a real polynomial of degree
$\gamma + \mathrm{N} + 1$ which vanishes in $z=0$, 
while $\mathrm{N} = (D-4)/2$
and $\gamma > D/2$ is integer\footnote{ 
$\gamma_E=0.577216$ is the Euler's constant, and  
\bee
\Gamma(b,z)=\int_{z}^{\infty}t^{b-1}e^{-t} dt
\eee 
is the incomplete gamma function \cite{M1}.}.
The exponential factor $e^{{\mathrm H}(z)}$ is crucial 
to make the theory super-renormalizable or finite 
at quantum level \cite{Krasnikov, Tomboulis, M1, BM, M2, M3, M4}. 

Let us expand on the behaviour of $\mathrm{H}(z)$ 
for small values of $z$:
\be
&& \hspace{-0.5cm} 
\mathrm{H}(z) = \sum_{n =1}^{\infty} \, 
\frac{p_{\gamma +\mathrm{N}+1}(z)^{2 n}}{2n \, ( -1 )^{n-1} \, n!}
\nonumber \\
&& \hspace{-0.5cm}
= \frac{1}{2} \left[ \gamma_E + 
\Gamma \left(0, p_{\gamma+\mathrm{N}+1}^{2}(z) \right)  
+ \log \left( p^2_{\gamma+ \mathrm{N}+1}(z) \right) \right] ,
\nonumber \\
&& \hspace{-0.5cm} \mbox{for} \,\,\, 
{\rm Re}( p_{\gamma+\mathrm{N}+1}^{2}(z) ) > 0 \,.
\label{HD}
\ee 
For the most simple choice $p_{\gamma+\mathrm{N}+1}(z) 
= z^{\gamma +\mathrm{N}+ 1}$, $\mathrm{H}(z)$ simplifies to
\be
&& \hspace{-0.5cm} 
\mathrm{H}(z) = \frac{1}{2} \left[ \gamma_E 
+ \Gamma \left(0, z^{2 \gamma +2 \mathrm{N}+2} \right) + \log (z^{2\gamma 
+2 \mathrm{N}+2}) \right] \, ,
\nonumber \\
&& \hspace{-0.5cm} 
{\rm Re}(z^{2 \gamma +2 \mathrm{N}+2}) > 0 \,  , \nonumber \\
&& \hspace{-0.5cm} 
\mathrm{H}(z) = \frac{ z^{2 \gamma + 2 \mathrm{N}+2}}{2} 
-\frac{ z^{4 \gamma + 4 \mathrm{N}+ 4}}{8} 
+ \dots \,\,\, {\rm for} \,\, z \approx 0  \, . 
\label{H0}
\ee
In particular $\lim_{z\rightarrow 0} {\mathrm H}(z) =0$.
We will expand more about the limit of large $z$ 
in Sec.~\ref{RENO}, where we will explicitly show 
the power counting renormalizability of the theory.

\subsection{Propagator}

In this section we calculate the two point function of 
the gravitational fluctuation around the flat space-time. 
For this purpose we split the $g_{\mu \nu}$ into 
the flat Minkowski metric $\eta_{\mu \nu}$
and the fluctuation $h_{\mu \nu}$, as
\bee 
g_{\mu \nu} =  \eta_{\mu \nu} + \kappa h_{\mu \nu}\,.
\eee
Writing the action (\ref{action}) in the form $S=\int d^D x\,\mathcal{L}$, 
the Lagrangian $\mathcal{L}$ can be expanded 
to second order in the graviton fluctuation \cite{HigherDG}
\be
&& \hspace{-0.4cm} \mathcal{L}_{\rm lin} = 
- \frac{1}{2} [ h^{\mu \nu } \Box h_{\mu \nu} + A_{\nu}^2 
+ (A_{\nu} - \phi_{, \nu})^2 ] \nonumber \\
&&  \hspace{-0.6cm} 
+ \frac{1}{4} \Big[ \frac{\kappa^2}{2} \Box h_{\mu\nu}  \beta( \Box) \Box h^{\mu \nu} 
- \frac{\kappa^2}{2} A^{\mu}_{, \mu}  \beta( \Box) A^{\nu}_{, \nu} 
 \nonumber \\
&&  \hspace{-0.6cm} - \frac{\kappa^2}{2} F^{\mu\nu}  \beta( \Box) F_{\mu \nu} 
+ \frac{\kappa^2}{2} (A^{\mu}_{, \mu} - \Box \phi) 
\beta( \Box) (A^{\nu}_{, \nu} - \Box \phi)
\nonumber \\
&& \hspace{-0.6cm} 
+ 2 \kappa^2 \left(A^{\mu}_{, \mu}  - \Box \phi \right) 
\alpha( \Box) (A^{\nu}_{, \nu} - \Box \phi ) \Big]  \,,
\label{quadratic2a} 
\ee
where $A^{\mu} = h^{\mu \nu}_{\,\,\,\, , \nu}$, $\phi = h$ (the trace of $h_{\mu \nu}$), 
$F_{\mu\nu} = A_{\mu , \nu} - A_{\nu, \mu}$ and 
the functionals of the D'Alembertian operator 
$\alpha(\Box), \beta (\Box)$ are defined by 
\be
&& \alpha(\Box)  :=  2  \sum_{n = 0}^{\mathrm{N} } 
a_n ( - \Box_{M})^n + 2 h_0(- \Box_{M}) , \nonumber \\
&& 
\beta(\Box)  :=  2  \sum_{n = 0}^{\mathrm{N}} 
b_n ( - \Box_{M})^n + 2 h_2(- \Box_{M})\,.
\label{alphabeta}
\ee

The d'Alembertian operator in Eq.~(\ref{quadratic2a}) 
must be conceived on the flat space-time. 
The linearized Lagrangian (\ref{quadratic2a}) is invariant under infinitesimal 
coordinate transformations $x^{\mu} \rightarrow x^{\mu} + \kappa \,  \xi^{\mu}(x)$, 
where $\xi^{\mu}(x)$ is an infinitesimal vector field of dimensions 
$[\xi(x)] = [{\rm mass}]^{(D-4)/2}$. 
Under this shift the graviton field is transformed as 
$h_{\mu \nu} \rightarrow h_{\mu \nu} - \xi_{\mu,\nu} - \xi_{\nu,\mu}$.
The presence of this local gauge invariance 
requires for a gauge-fixing term
to be added to the linearized Lagrangian (\ref{quadratic2a}). 
Hence, if we choose the usual harmonic gauge 
($\partial^{\mu } h_{\mu \nu} = 0$) \cite{Tomboulis, Stelle} 
\bee
\mathcal{L}_{\rm GF} =   
\xi^{-1} 
\partial^{\mu} h_{\mu \nu}\, V^{-1}(-\Box_{M}) \, 
\partial_{\rho} h^{\rho \nu}\,,
\label{GF2}
\eee
the linearized gauge-fixed Lagrangian reads 
\be
\mathcal{L}_{\rm lin} + \mathcal{L}_{\rm GF} = 
\frac{1}{2} h^{\mu \nu} \mathcal{O}_{\mu\nu, \rho \sigma} 
\, h^{\rho \sigma },
\label{O}
\ee
where the operator $\mathcal{O}$ is made of two terms, 
one coming from the linearized Lagrangian 
(\ref{quadratic2a}) and the other from the gauge-fixing term (\ref{GF2}).

Inverting the operator $\mathcal{O}$ \cite{HigherDG}, we find 
the following two-point function in the momentum space (with 
the wave number $k$), 
\be
\hspace{-0.5cm}  
\mathcal{O}^{-1} &=& \frac{\xi (2P^{(1)} + \bar{P}^{(0)} ) }{2 k^2 \, V^{-1}( k^2/M^2)} 
+ \frac{P^{(2)}}{k^2 \Big(1 + \frac{k^2 \kappa^2 \beta(k^2)}{4} \Big)} 
\nonumber \\
\hspace{-0.5cm}
& &- \frac{P^{(0)}}{2 k^2 \Big( \frac{D-2}{2} - k^2 \frac{D \beta(k^2) 
\kappa^2/4 + (D-1) \alpha(k^2) \kappa^2}{2} \Big) } \,. 
\label{propagator}
\ee
where we omitted the tensorial indices for $\mathcal{O}^{-1}$. 
The operators \{$P^{(2)}$, $P^{(1)}$, 
$P^{(0)}$, $\bar{P}^{(0)}$\}, which project out the 
spin-2, spin-1, and two spin-0 parts of a massive tensor field, 
are defined by \cite{HigherDG}\label{proje2}
\be
 && \hspace{-0.2cm} 
 P^{(2)}_{\mu \nu, \rho \sigma}(k) = \frac{1}{2} 
 ( \theta_{\mu \rho} \theta_{\nu \sigma} +
 \theta_{\mu \sigma} \theta_{\nu \rho} ) 
 - \frac{1}{D-1} \theta_{\mu \nu} \theta_{\rho \sigma} \, , 
 \nonumber 
 \\ 
 \nonumber \\
 && \hspace{-0.2cm}
   P^{(1)}_{\mu \nu, \rho \sigma}(k) = \frac{1}{2} \left( \theta_{\mu \rho} \omega_{\nu \sigma} +
 \theta_{\mu \sigma} \omega_{\nu \rho}  + 
 \theta_{\nu \rho} \omega_{\mu \sigma}  +
  \theta_{\nu \sigma} \omega_{\mu \rho}  \right) \, , \nonumber   \\
   &&
  \hspace{-0.2cm} 
 P^{(0)} _{\mu\nu, \rho\sigma} (k) = \frac{1}{D-1}  \theta_{\mu \nu} \theta_{\rho \sigma}  \, , \,\,\,\, \,\,
 \bar{P}^{(0)} _{\mu\nu, \rho\sigma} (k) =  \omega_{\mu \nu} \omega_{\rho \sigma} \, ,  
 \ee
where $\omega_{\mu \nu}=k_{\mu} k_{\nu}/k^2$ and 
$\theta_{\mu \nu}=\eta_{\mu \nu} - k_{\mu} k_{\nu}/k^2$.
These correspond to a complete set of projection operators for 
symmetric rank-two tensors.
The functions $\alpha(k^2)$ and $\beta(k^2)$ are achieved by replacing 
$\Box \rightarrow -k^2$ in the definitions (\ref{alphabeta}).

By looking at the last two gauge-invariant terms in Eq.~(\ref{propagator}), 
we deem convenient to introduce the following definitions, 
\be
&& \hspace{-0.4cm}
\bar{h}_2(z)  =  1 + \frac{\kappa^2 M^2}{2}  
z \sum_{n=0}^{\mathrm{N}} b_n z^n + \frac{\kappa^2 M^2}{2}
z \, h_2(z) \, , \label{barh2h0} \\
&& \hspace{-0.4cm}
\frac{D-2}{2}\,\bar{h}_0(z) =  \frac{D-2}{2} 
- \frac{\kappa^2 M^2 D}{4}  z 
\left[\sum_{n=0}^{\mathrm{N}} b_n z^n + h_2(z) \right] 
\nonumber \\
&& \hspace{1.7cm}
-\kappa^2 M^2 (D - 1)  z \left[\sum_{n=0}^{\mathrm{N}}
a_n z^n + h_0(z) \right].
\label{barh2h0d}
\ee
Through these definitions, the gauge-invariant part 
of the propagator greatly simplifies to
\be
\hspace{-0.1cm} 
\mathcal{O}^{-1}
= \frac{1}{k^2}
\left[ \frac{P^{(2)}}{\bar{h}_2} 
- \frac{P^{(0)}}{(D-2) \bar{h}_0 } \right]\,.
\label{propgauge0}
\ee
\subsection{Power counting super-renormalizability}
\label{RENO}

The main properties of the entire function $e^{{\mathrm H}(z)}$ 
useful to show the super-renormalizability of the theory are the following,
\be
&& \hspace{-0.5cm} \lim_{z \rightarrow +\infty} e^{{\mathrm H}(z)} 
= e^{\frac{\gamma_E}{2}} \, |z|^{\gamma + \mathrm{N} +1} \,\,\,\, 
\,\,\,\, {\rm and} \nonumber \\
&& \hspace{-0.5cm} 
 \lim_{z \rightarrow +\infty} 
\left(\frac{ e^{{\mathrm H}(z)}}{e^{\frac{\gamma_E}{2}} 
|z|^{\gamma + \mathrm{N} +1} } - 1 \right) z^n = 0
\,\,\,\, \forall \, n \in \mathbb{N}\, , 
\label{property}
\ee
where we assumed $p_{\gamma + {\mathrm N} + 1}(z) 
= z^{\gamma + {\mathrm N} + 1}$.
The first limit tells us what is the leading behaviour in the ultraviolet regime, 
while the second limit confirms that the next to the leading order goes to zero 
faster then any polynomial. 

Let us then examine the ultraviolet behavior of the theory at quantum level.
According to the property (\ref{property}), the propagator 
and the leading $n$-graviton interaction vertex have the same scaling
in the high-energy regime [see Eqs.~(\ref{hzD}), (\ref{Vdef}), 
(\ref{barh2h0}), (\ref{propgauge0}), and (\ref{property})]:
\be
&& \hspace{-1cm} 
{\rm propagator}: \, \,\, \,
\mathcal{O}^{-1} \sim \frac{1}{k^{2 \gamma +2 \mathrm{N} +4}} \, , 
\label{propK} \\
&& \,\, \hspace{-1.1cm}
{\rm{vertex}} : \, \,\, \, 
{\mathcal L}^{(n)} \sim  h^n \, \Box_{\eta} h \,
h_i( - \Box_{M}) \,\, \Box_{\eta} h 
\nonumber  \\
&& \hspace{-1cm}
\,\, \rightarrow \,\, h^n \, \Box_{\eta} h 
\,  ( \Box_{\eta} + h^m \, \partial h
 \partial )^{\gamma + \mathrm{N} } \, 
\Box_{\eta} h  \sim
k^{2 \gamma +2 \mathrm{N}+4}\,.  
\label{intera3}
\ee

In Eq.~(\ref{intera3}) the indices for the graviton fluctuation $h_{\mu \nu}$ 
are omitted and $h_i( - \Box_{M})$ is one of the functions in Eq.~(\ref{hzD}). 
From Eqs.~(\ref{propK}) and (\ref{intera3}), 
the upper bound to the superficial degree of divergence is 
\be
\omega &=& 
D L - (2 \gamma + 2 \mathrm{N}+ 4) I 
+ (2 \gamma + 2 \mathrm{N} + 4) V 
\nonumber \\
&=& D - 2 \gamma  (L - 1).
\label{diverE}
\ee
In Eq.~(\ref{diverE})  we used the topological relation between vertexes $V$, 
internal lines $I$ and number of loops $L$: $I = V + L -1$, 
as well as $D=2 \mathrm{N}+4$.
Thus, if $\gamma > D/2$, then only 1-loop divergences survive 
and the theory is super-renormalizable. 
Only a finite number of constants is renormalized in the action (\ref{action}), i.e. 
$\kappa$, $\bar{\lambda}$, $a_n$, $b_n$ together with the finite number 
of couplings that multiply the operators $O(R^3)$ in 
the last line of Eq.~(\ref{action}).

We now assume that the theory is renormalized at some scale $\mu_0$. 
Therefore, if we set
\be
\tilde{a}_n = a_n(\mu_0) \,,
\,\,\,\, \tilde{b}_n = b_n(\mu_0) \,,
\label{betaalphaD}
\ee
in Eq.~(\ref{hzD}), the functions (\ref{barh2h0}) and 
(\ref{barh2h0d}) reduce to
\be
\bar{h}_2 = \bar{h}_0 = V^{-1}(z) = 
\frac{\square + m^2}{\square} 
e^{{\mathrm H}(z)}\,.
\ee
Thus, in the momentum space, only a pole at $k^2 = m^2$ occurs 
in the bare propagator and Eq.~(\ref{propgauge0}) reads 
\be
\hspace{-0.25cm}
\mathcal{O}^{-1}\! =  
\frac{e^{-{\mathrm H}  
\left( \! \frac{k^2}{M^2} \! \right)}}{k^2-m^2} \!
\left[ 
P^{(2)}- \frac{P^{(0)}}{D-2} +
 \xi  \left( \! P^{(1)}+\frac{\bar{P}^{(0)}}{2}  \right) 
\right] \!  .
\label{propgauge2b}
\ee
The tensorial structure of Eq.~(\ref{propgauge2b}) is the same as that of 
the massless graviton and the only difference appears 
in an overall factor $1/(k^2 - m^2)$. 
If we take the limit $m\rightarrow 0$, the massive graviton 
propagator reduces smoothly to the massless one 
and hence there is no vDVZ discontinuity. 

Assuming the renormalization group invariant condition (\ref{betaalphaD}), 
missing the $O(R^3)$ operators in the action (\ref{action}), and  
setting $\bar{\lambda}$ to zero, the non-local Lagrangian in 
a $D$ dimensional space-time greatly simplifies to 
\be
\hspace{-0.15cm} 
\mathcal{L} \! = \!
\frac{2}{\kappa^{2}} \sqrt{|g|} 
\left[ R - G_{\mu \nu} 
 \frac{ (\Box + m^2) 
e^{{\mathrm H}(-\Box_{M}) } 
-\Box }{\Box^2} 
R^{\mu \nu} \right]\! .
\label{NLL}
\ee
On using the function $\alpha(\Box)=2(V(\Box)^{-1} -1)/(\kappa^2 \Box)$, 
the Lagrangian (\ref{NLL}) can be expressed as
\bee
\mathcal{L} = 
\sqrt{|g|} \left[ \frac{2}{\kappa^2}R+\frac12 R \, \alpha (\Box) R
-R_{\mu \nu} \alpha(\Box) R^{\mu \nu} \right]\,.
\label{NLL2}
\eee
If we are interested only in the infrared modifications of gravity, 
we can fix ${\mathrm H}(- \Box_M)=0$. 
This condition restricts our class of theories to 
the non-local massive gravity.  

\subsection{Unitarity}

We now present a systematic study of the tree-level unitarity \cite{HigherDG}. 
A general theory is well defined if  ``tachyons" and ``ghosts" are absent, 
in which case the corresponding propagator has only first poles at $k^2 - m^2 =0$ 
with real masses (no tachyons) and with positive
residues (no ghosts). Therefore, to test the tree-level unitarity, 
we couple the propagator to external conserved 
stress-energy tensors, $T^{\mu \nu}$, and we examine 
the amplitude at the poles \cite{VanNieuwenhuizen:1973fi}.
When we introduce the most general source, 
the linearized action (\ref{O}) is replaced by 
\bee
\mathcal{L}_{\rm lin} + \mathcal{L}_{\rm GF} 
\,\, \rightarrow \,\, 
 \frac{1}{2} h^{\mu \nu} \mathcal{O}_{\mu \nu , \rho \sigma} h^{\rho \sigma} 
- g h_{\mu \nu} T^{\mu \nu}\,,
\label{LGM}
\eee
where $g$ is a coupling constant. 
The transition amplitude in the momentum space is 
\bee
\mathcal{A} = g^2 \, T^{\mu \nu} \, 
\mathcal{O}^{-1}_{\mu \nu , \rho \sigma} \, T^{\rho \sigma} \,.
\label{ampli1}
\eee
Since the stress-energy tensor is conserved, only the projectors 
$P^{(2)}$ and $P^{(0)}$ will give non-zero contributions to the amplitude.

In order to make the analysis more explicit, we expand the sources using the 
following set of independent vectors in the momentum 
space \cite{VanNieuwenhuizen:1973fi, Veltman:1975vx, HigherDG, Accioly:2011nf}:
\be
&& k^{\mu} = (k^0, \vec{k}) \, , \,\, \tilde{k}^{\mu} = (k^0, - \vec{k}) \, , \nonumber \\ 
&& \epsilon^{\mu}_i = (0, \vec{\epsilon}_i) \, , \,\, i =1, \dots , D-2 \, ,
\ee
where $\vec{\epsilon}_i$ are unit vectors orthogonal 
to each other and to $\vec{k}$.
The symmetric stress-energy tensor reads 
\be
&& T^{\mu\nu} = a k^{\mu} k^{\nu} + b \tilde{k}^{\mu} \tilde{k}^{\nu} 
+ c^{i j} \epsilon_i^{(\mu} \epsilon_j^{\nu)} + d \, k^{(\mu} \tilde{k}^{\nu)} \nonumber \\
&& \hspace{1cm}
+ e^i k^{(\mu} \epsilon_i^{\nu)} + f^i \tilde{k}^{(\mu} \epsilon_i^{\nu)} \,,
\ee
where we introduced the notation $X_{(\mu} Y_{\nu)} \equiv (X_{\mu} Y_{\nu}+Y_{\mu} X_{\nu})/2$. 
The conditions $k_{\mu} T^{\mu \nu} =0$ and $k_{\mu} k_{\nu}T^{\mu \nu} =0$ place 
the following constraints on the coefficients $a,b,d, e^i, f^i$ \cite{HigherDG}:
\be
&& a k^2 + (k_0^2 + \vec{k}^2) d/2=0\,, \label{1} \\
&& b (k_0^2 + \vec{k}^2) + d k^2/2 =0\,, \label{2} \\
&& e^i k^2 + f^i (k_0^2 + \vec{k}^2) = 0\,, \label{3} \\
&& a k^4 +b (k_0^2 + \vec{k}^2)^2 + d \, 
k^2 (k_0^2 + \vec{k}^2) = 0\,, \label{4}
\ee
where $k^2:=k_0^2-\vec{k}^2$.
The conditions (\ref{1}) and (\ref{2}) imply 
\bee
a (k^2)^2 = b (k_0^2 + \vec{k}^2)^2 \,\, \Longrightarrow \,\, a \geqslant b \,,
\label{cond1} 
\eee
while the condition (\ref{3}) leads to
\bee
(e^i)^2 = (f^i)^2 \left(\frac{k_0^2 + \vec{k}^2}{k^2}\right)^2 \,\, 
\Longrightarrow \,\, (e^i)^2  \geqslant (f^i)^2\,.
\label{cond2} 
\eee

Introducing the spin-projectors and the conservation of 
the stress-energy tensor $k_{\mu} T^{\mu \nu} = 0$ in Eq.~(\ref{ampli1}), 
the amplitude results 
\be
\mathcal{A} = g^2 \left(  T_{\mu \nu} T^{\mu \nu} - \frac{T^2}{D-2} \right) 
\frac{e^{- {\mathrm H}(k^2/M^2)}}{k^2 - m^2} \,,
\label{ampli2}
\ee
where $T:=\eta^{\mu \nu} T_{\mu \nu}$.
The residue at the pole $k^2 = m^2$ reads
\be
&& \hspace{-0.4cm}
{\rm Res} \, \mathcal{A}  \Big|_{k^2 = m^2} \!\!\!\!\!\! 
= g^2 \bigg\{  [(a-b)k^2]^2 + (c^{i j})^2 + \frac{k^2}{2} [ (e^i)^2 - (f^i)^2] 
\nonumber \\
&& \hspace{-0.4cm}
- \frac{1}{D-2} [ ( b-a) k^2 - c^{ii}]^2 \bigg\}
e^{- {\mathrm H} \left( \frac{k^2}{M^2} \right)}
\bigg|_{k^2 = m^2} \label{amp1}\\
&& \hspace{-0.4cm}
= g^2 e^{- {\mathrm H} \left( \frac{m^2}{M^2} \right)}
\bigg\{ \frac{D-3}{D-2} [(a-b) m^2]^2 
+ \left[ (c^{i j})^2 - \frac{(c^{ii})^2}{D-2} \right] 
\nonumber \\
&& \hspace{-0.4cm}
+\frac{m^2}{2} [ (e^i)^2 - (f^i)^2]
- \frac{2}{D-2}(a-b) m^2 c^{ii} \bigg\}. 
\label{amplitudem2}
\ee
If we assume the stress-tensor to satisfy the usual energy condition, 
then the following inequality follows
\bee
T = (b-a) k^2 - c^{ii} \geqslant 0 \,\, \Longrightarrow \,\, 
c^{ii} \leqslant 0 \,.
\label{trace}
\eee
Using the conditions (\ref{cond1}), (\ref{cond2}), and (\ref{trace}) 
in Eq.~(\ref{amplitudem2}), we find that 
\bee
{\rm Res} \, \mathcal{A}  \Big|_{k^2 = m^2} \geqslant 0\,, 
\eee
for $D \geq 3$.
This shows that the theory is unitary at tree level around 
the Minkowski background.
As we see in Eq.~(\ref{amp1}) the contribution to the 
residue from the spin-0 operator $P^{(0)}$ is negative, 
but the spin-2 operator $P^{(2)}$ provides a dominant 
contribution with a positive sign of 
${\rm Res} \, \mathcal{A}  \big|_{k^2 = m^2}$. 
Hence the presence of the spin-2 mode is crucial 
to make the theory unitary.

\subsection{Equations of motion}

Let us derive the equations of motion up to curvature squared 
operators $O(R^2)$ and total derivative 
terms \cite{barvy, barvy2, barvy3, barvy4,Arkani}.
The action of our theory is $S=\int d^D x {\cal L}$, where 
the Lagrangian is given by Eq.~(\ref{NLL}). 
The variation of this action reads
\be
&& \hspace{-0.35cm} \delta S \! =  \! 
\frac{2}{\kappa^2} \!\!
\int \! \! d^D x \left[\delta (\sqrt{|g|} R) 
-\delta \! \left( \sqrt{|g|}G^{\mu \nu} \frac{V^{-1} -1}{\Box} R_{\mu \nu} \!
\right) \right] \nonumber \\
&& \hspace{-0.35cm}
= \frac{2}{\kappa^2} 
\int \! \! d^D x \sqrt{|g|}\Big[ G_{\mu \nu} \delta g^{\mu \nu} 
- 2 G^{\mu \nu}  \frac{ V^{-1} -1}{\Box} \delta R_{\mu \nu}  + \dots \Big]  
\nonumber \\
&&  \hspace{-0.35cm} =\frac{2}{\kappa^2} 
\int \! \! d^D x \! \sqrt{|g|} \left[ V^{-1} \, G_{\mu \nu} \, \delta g^{\mu \nu} +
O(R^2) \right]\,,
\label{variation}
\ee
where we omitted the argument $-\Box_{M}$ of the form factor $V^{-1}$. 
We also used the relations 
$\nabla_{\mu} g_{\rho \sigma} =0$, $\nabla^{\mu} G_{\mu \nu} =0$, and
\be
\hspace{-0.4cm}
\delta R_{\mu \nu} &=&
- \frac{1}{2} g_{\mu \alpha} g_{\nu \beta} \Box \delta g^{\alpha \beta}  \\
\hspace{-0.6cm} & &
- \frac{1}{2} \big[ \nabla^{\beta} \nabla_{\mu} \delta g_{\beta \nu} 
+ \nabla^{\beta} \nabla_{\nu} \delta g_{\beta \mu} - \nabla_{\mu} \nabla_{\nu} 
\delta g_{\alpha}^{\alpha} \big]. \nonumber
\ee
The action is manifestly covariant in general. 
Hence its variational derivative (the left hand side of the modified Einstein equations) 
exactly satisfies the Bianchi identity
\be
\nabla^{\mu} \frac{\delta S}{\delta g^{\mu \nu}} = \sqrt{|g|} \, \nabla_{\mu}
\left[ V^{-1}(\Box) \, G_{\mu \nu} + O(R_{\mu\nu}^2)  \right] =0 \, . 
\ee
Taking into account the energy-momentum tensor $T_{\mu \nu}$, 
the equation of motion at the quadratic order of curvatures reads 
\be
V^{-1}(\Box) \, G_{\mu \nu} + O(R_{\mu \nu}^2) 
= 8 \pi G T_{\mu \nu} \,. 
\label{EMT}
\ee 

Except for the very high-energy regime the quadratic curvature 
terms should not be important in Eq.~(\ref{EMT}).
Neglecting the $O(R_{\mu \nu}^2)$ terms and setting 
$e^{{\mathrm H}(-\Box_{M})}=1$ in Eq.~(\ref{EMT}), it follows that 
\be
G_{\mu \nu}+\frac{m^2}{\Box} G_{\mu \nu} 
\simeq 8\pi G \, T_{\mu \nu}\,,
\label{EMT2}
\ee 
which is the same equation as that studied in Ref.~\cite{Maggiore} 
in the context of non-local massive gravity with the graviton mass $m$.
 
If we apply Eq.~(\ref{EMT2}) to cosmology, 
the d'Alembertian is of the order of $\Box \sim d^2/dt^2 \sim \omega^2$, 
where $\omega$ is the characteristic frequency of a corresponding 
physical quantity.
Provided $\omega \gg m$ the term 
$m^2 \Box^{-1} G_{\mu \nu}$ in Eq.~(\ref{EMT2}) is 
suppressed relative to $G_{\mu \nu}$, so that 
the Einstein equation 
$G_{\mu \nu} \simeq 8\pi G T_{\mu \nu}$ is recovered.
In order to realize the standard radiation and matter eras, 
it is expected that $m$ should not be larger than $H_0$.
At the late cosmological epoch, the effect of the non-local term 
$m^2 \Box^{-1} G_{\mu \nu}$ can be important to modify  
the dynamics of the system.

If we take the derivative of Eq.~(\ref{EMT2}) by exerting the 
operator $\Box$, it follows that  
\bee
\left( \Box+m^2 \right)G_{\mu \nu}
= 8 \pi G\,\Box T_{\mu \nu} \,.
\label{eqmR}
\eee
This equation is invariant under the symmetry 
\bee
T_{\mu \nu} \rightarrow T_{\mu \nu} 
+ ({\rm constant})\,g_{\mu\nu}\,,
\eee
which realizes the Afshordi-Smolin idea \cite{smolin} for 
the degravitation of the cosmological constant. 
Equation (\ref{eqmR}) does not admit exact de Sitter 
solutions.
There exist de-Sitter solutions characterized by 
$G_{\mu \nu}^{\rm dS}=8\pi G\,\rho_{\Lambda}^{\rm eff}g_{\mu \nu}$
for the modified model in which 
the operator $\Box$ in Eq.~(\ref{eqmR}) 
is replaced by $\Box+\mu^2$, where 
$\mu$ is a small mass scale \cite{Maggiore}. 
If the energy-momentum tensor on the right hand side 
of Eq.~(\ref{eqmR}) is given by 
$T_{\mu \nu}^{(\Lambda)}=\rho_{\Lambda}g_{\mu \nu}$, 
we obtain
the effective cosmological constant 
$\rho_{\Lambda}^{\rm eff}=\rho_{\Lambda}\,\mu^2/(m^2+\mu^2)$.
For $\mu$ much smaller than $m$, it follows that 
$\rho_{\Lambda}^{\rm eff} \ll \rho_{\Lambda}$.
In the limit $\mu \to 0$, the effective cosmological constant 
disappears completely.

The crucial point for the above degravitation of
$\rho_{\Lambda}$ is that both $\Box G_{\mu \nu}^{\rm dS}$ and 
and $\Box T_{\mu \nu}^{(\Lambda)}$ vanish at de Sitter solutions.
For the background in which the matter density $\rho$ varies 
(such as the radiation and matter eras), the two d'Alembertians 
in Eq.~(\ref{eqmR}) give rise to the contributions of the 
order of $\omega^2$. In other words, the above degravitation of 
$\rho_{\Lambda}$ should occur at the late cosmological epoch 
in which $\omega$ drops below $\mu$ \cite{Maggiore}.

A detailed analysis given in Sec.~\ref{cosmosec} shows that, 
even for $\rho_{\Lambda}=0$ and $\mu=0$, a late-time cosmic
acceleration occurs on the flat FLRW background.
This comes from the peculiar evolution of the term 
$m^2 \Box^{-1}G_{\mu \nu}$ in Eq.~(\ref{EMT2}), by 
which the equation of state smaller than $-1$ 
can be realized. Even in the presence of the cosmological 
constant, the non-local term eventually dominates 
over $\rho_{\Lambda}$ at the late cosmological epoch.
In the following we focus on the theory based on the field 
equation (\ref{EMT2}), i.e., $\mu=0$.

\section{Cosmological dynamics}
\label{cosmosec}

We study the cosmological dynamics on the four-dimensional 
flat FLRW background characterized by the line element 
$ds^2=-dt^2+a^2(t)(dx^2+dy^2+dz^2)$, where $a(t)$ 
is the scale factor with the cosmic time $t$.
Since we ignore the $O(R_{\mu \nu}^2)$ terms and set 
${\mathrm H}(-\Box_{M})=0$ in Eq.~(\ref{EMT}), 
our analysis can be valid in the 
low-energy regime much below the Planck scale.

We introduce a tensor $S_{\mu \nu}$ satisfying 
the relation 
\bee
\Box S_{\mu \nu} =G_{\mu \nu}\,,
\label{Smure}
\eee
by which the second term on the left hand side of Eq.~(\ref{EMT2}) 
can be written as $m^2 \Box^{-1} G_{\mu \nu}=m^2 S_{\mu \nu}$.
In order to respect the continuity equation 
$\nabla^{\mu} T_{\mu \nu}=0$ of matter, we take 
the transverse part $S_{\mu \nu}^{\rm T}$ of the symmetric tensor 
$S_{\mu \nu}$, that is, $\nabla^{\mu} S_{\mu \nu}^{\rm T}=0$.
Then, Eq.~(\ref{EMT2}) can be written as 
\bee
G_{\mu \nu}+m^2 S_{\mu \nu}^{\rm T}=
8\pi G\,T_{\mu \nu}\,.
\label{Ein}
\eee
We use the fact that $S_{\mu \nu}$ can 
decomposed as \cite{Porrati,Maggiore}
\be
S_{\mu \nu}=S_{\mu \nu}^{\rm T}+(\nabla_{\mu} S_{\nu}
+\nabla_{\nu} S_{\mu})/2\,,
\label{Smunu}
\ee
where the vector $S_{\mu}$ has the time-component 
$S_0$ alone in the FLRW background, i.e., 
$S_i=0$ ($i=1,2,3$).

{}From Eq.~(\ref{Smunu}) we have 
\bee
(S^0_0)^{\rm T}=u-\dot{S}_0\,,\qquad
(S^i_i)^{\rm T}=v-3HS_0\,,
\eee
where $u \equiv S^0_0$ and $v \equiv S^i_i$, and 
a dot represents a derivative with respect to $t$.
In the presence of the matter energy-momentum tensor 
$T_{\mu \nu}=(\rho, a^2P \delta_{ij})$, the (00) and 
$(ii)$ components of Eq.~(\ref{Ein}) are
\be
& & 3H^2+m^2 (u-\dot{S}_0)=8\pi G \rho\,,\label{Heq}\\
& & 2\dot{H}+3H^2+\frac{m^2}{3} 
(v-3HS_0)=-8\pi G P\,,
\ee
where $H=\dot{a}/a$.

Taking the divergence of Eq.~(\ref{Smunu}), it follows that 
$2\nabla^{\mu} S_{\mu \nu}=
\nabla^{\mu} (\nabla_{\mu} S_{\nu}+\nabla_{\nu} S_{\mu})$.
{}From the $\nu=0$ component of this equation we obtain
\bee
S_0=\frac{1}{\partial_0^2+3H\partial_0-3H^2}
(\dot{u}+3Hu-Hv)\,.
\eee
The (00) and $(ii)$ components of Eq.~(\ref{Smure}) give
\be
& & \ddot{u}+3H \dot{u}-6H^2u+2H^2 v=3H^2\,,\\
& & \ddot{v}+3H \dot{v}+6H^2u-2H^2v=6\dot{H}+9H^2\,,
\label{ddotv}
\ee
which can be decoupled each other by 
defining 
\be 
U \equiv u+v \,\, \,\, \mbox{and} \,\, \,\,  V \equiv u-\frac{v}{3} \, .
\ee
In summary we get the following set of equations 
from Eqs.~(\ref{Heq})-(\ref{ddotv}):
\be
& & \hspace{-0.5cm}
3H^2+\frac{m^2}{4}(U+3V-4\dot{S}_0)=8\pi G \rho\,,\label{be1}\\
& & \hspace{-0.5cm}
2\dot{H}+3H^2+\frac{m^2}{4} (U-V-4HS_0)=-8\pi G P,\label{be2}\\
& & \hspace{-0.5cm}
\ddot{S}_0+3H\dot{S}_0-3H^2 S_0=\frac14
(\dot{U}+3\dot{V}+12HV)\,,\label{be3}\\
& &\hspace{-0.5cm}
 \ddot{U}+3H \dot{U}=6(\dot{H}+2H^2)\,,\\
& & \hspace{-0.5cm}
\ddot{V}+3H \dot{V}-8H^2 V=-2\dot{H}\,.
\label{ddotv2}
\ee

From Eqs.~(\ref{be1})-(\ref{be3}) one can show that the 
continuity equation $\dot{\rho}+3H(\rho+P)=0$ holds.
For the matter component we take into account 
radiation (density $\rho_r$, pressure $P_r=\rho_r/3$),
non-relativistic matter (density $\rho_m$, pressure $P_m=0$), 
and the cosmological constant (density $\rho_\Lambda$, 
pressure $P_\Lambda=-\rho_\Lambda$), i.e., 
$\rho=\rho_r+\rho_m+\rho_{\Lambda}$ and 
$P=\rho_r/3-\rho_{\Lambda}$.
Each matter component obeys the continuity equation 
$\dot{\rho}_i+3H(\rho_i+P_i)=0$ ($i=r,m,\Lambda$).

In order to study the cosmological dynamics of the above system, 
it is convenient to introduce the following dimensionless variables
\be
&& \hspace{-0.7cm} 
S=HS_0,\quad
\Omega_r=\frac{8\pi G \rho_r}{3H^2},\quad
\Omega_{\Lambda}=\frac{8\pi G \rho_{\Lambda}}{3H^2},\nonumber \\
&&  \hspace{-0.7cm}
\Omega_{\rm NL}=\frac{m^2}{12H^2} 
\left( 4S'-4Sr_H-U-3V \right)\,,
\ee
where $r_H \equiv H'/H$, and a prime represents a derivative 
with respect to $N=\ln (a/a_i)$ ($a_i$ is the initial scale factor).
{}From Eq.~(\ref{be1}) it follows that 
\bee
\Omega_m \equiv \frac{8\pi G \rho_m}{3H^2}
=1-\Omega_r-\Omega_\Lambda-\Omega_{\rm NL}.
\eee
We define the density parameter of the dark energy 
component, as $\Omega_{\rm DE} 
\equiv \Omega_\Lambda+\Omega_{\rm NL}$. 
From Eqs.~(\ref{be1}) and (\ref{be2}) the density and 
the pressure of dark energy are given respectively by 
\be
&& \rho_{\rm DE}=\rho_{\Lambda} - \frac{m^2}{32\pi G} 
(U+3V-4\dot{S}_0)\,,
\nonumber \\
&& P_{\rm DE}=-\rho_{\Lambda} + \frac{m^2}{32\pi G} 
(U-V-4HS_0) \,.
\ee 
Then, the dark energy equation of state 
$w_{\rm DE}=P_{\rm DE}/\rho_{\rm DE}$
can be expressed as
\bee
w_{\rm DE}=-\frac{\Omega_{\Lambda}-(U-V-4S)m^2/(12H^2)}
{\Omega_{\Lambda}-(U+3V-4S'+4Sr_H)m^2/(12H^2)}\,.
\label{wde}
\eee

From Eq.~(\ref{be2}) the quantity $r_H=H'/H$ obeys
\bee
r_H=-\frac32-\frac12 \Omega_r+\frac32 \Omega_{\Lambda}
-\frac{m^2}{8H^2} (U-V-4S)\,,
\label{rH}
\eee
by which the effective equation of state of the Universe 
is known as $w_{\rm eff}=-1-2r_H/3$.
On using Eqs.~(\ref{be2})-(\ref{ddotv2}) and the continuity
equation of each matter component, 
we obtain the following differential equations
\be
& &U''+(3+r_H)U'=6(2+r_H)\,,\label{auto1}\\
& &V''+(3+r_H)V'-8V=-2r_H\,,\label{auto2}\\
& & S''+(3-r_H)S'-(3+3r_H+r_H')S \nonumber \\
& & =\frac14 
\left( U'+3V'+12V \right)\,,\label{auto3}\\
& &\Omega_r'+(4+2r_H) \Omega_r=0\,,\\
& &\Omega_{\Lambda}'+2r_H \Omega_{\Lambda}=0\,.
\label{auto5}
\ee
In Eq.~(\ref{auto3}) the derivative of $r_H$ is given by 
\bee
r_H'=2\Omega_r-3r_H-2r_H^2
-\frac{m^2}{8H^2} (U'-V'-4S')\,.
\eee
\subsection{$\Omega_{\Lambda}=0$}
\label{nosec}

Let us first study the case in which the cosmological constant 
is absent ($\Omega_{\Lambda}=0$). 
We assume that $m$ is smaller than the today's Hubble 
parameter $H_0$, i.e., $m \lesssim H_0$.
During the radiation and matter dominated epochs
the last term in Eq.~(\ref{rH}) should be suppressed, 
so that $r_H \simeq -3/2-\Omega_r/2$ is nearly constant 
in each epoch. 
Integrating Eqs.~(\ref{auto1}) and (\ref{auto2}) 
for constant $r_H$ ($>-3$) and neglecting the 
decaying modes, we obtain
\be
U &=& c_1+\frac{6(2+r_H)N}{3+r_H}\,,\label{Uso}\\
V &=& c_2e^{\frac12 \left( -3-r_H+\sqrt{41+6r_H+r_H^2} \right)N}
+\frac14 r_H\,, \label{Vso}
\ee
where $c_1$ and $c_2$ are constants. 
During the radiation era ($r_H=-2$) these solutions 
reduce to $U=c_1$ and $V=c_2e^{(\sqrt{33}-1)N/2}-1/2$, 
while in the matter era ($r_H=-3/2$) we have 
$U=2N+c_1$ and $V=c_2 e^{(\sqrt{137}-3)N/4}-3/8$.
Since $V$ grows faster than $U$ due to the presence of 
the term $-8V$ in Eq.~(\ref{auto2}), it is a good approximation 
to neglect $U$ relative to $V$
in the regime $|V| \gg 1$.

The field $S$ is amplified by the force 
term on the right hand side of Eq.~(\ref{auto3}). 
Meanwhile the homogeneous solution of Eq.~(\ref{auto3}) 
decays for $r_H=-2$ and $-3/2$.
Then, for $|V| \gg 1$, the field $S$ grows as
\bee
S \simeq \frac{3(25+11r_H+5\sqrt{41+6r_H+r_H^2})}
{8(25-25r_H-6r_H^2)}V\,,
\label{Sso}
\eee
which behaves as $S \simeq (5\sqrt{33}+3)c_2e^{(\sqrt{33}-1)N/2}/136$
during the radiation era and 
$S \simeq 3(5\sqrt{137}+17)c_2e^{(\sqrt{137}-3)N/4}/784$
during the matter era.
{}From Eq.~(\ref{wde}) the dark energy equation of state 
reduces to $w_{\rm DE} \simeq (V+4S)/(3V-4S'+4Sr_H)$. 
Using the above solutions, we obtain 
\be
&& \hspace{-1.1cm}
w_{\rm DE} \simeq \nonumber \\
&&  \hspace{-1.1cm}
-\frac13
\frac{125-17r_H-12r_H^2+15\sqrt{41+6r_H+r_H^2}}
{15+11r_H-2r_H^2+(5-2r_H)\sqrt{41+6r_H+r_H^2}},
\label{wdeso}
\ee
from which 
$w_{\rm DE} \simeq -1.791$ in the radiation era and 
$w_{\rm DE} \simeq -1.725$ in the matter era.
This means that the dark component from the non-local mass
term comes into play at the late stage of cosmic expansion history. 

Indeed, there exists an asymptotic future solution characterized by 
$\Omega_{\rm NL}=1$ with constant $r_H$.
In this regime we have $r_H \simeq -3/2-m^2/(8H^2)(U-V-4S)$
in Eq.~(\ref{rH}). Meanwhile, if $r_H$ is constant,  
the mass term $m$ does not appear in Eqs.~(\ref{auto1})-(\ref{auto3}), 
so that the solutions (\ref{Uso})-(\ref{wdeso}) are valid, too. 
Since $w_{\rm DE}$ is equivalent to $w_{\rm eff}=-1-2r_H/3$ 
in the limit $\Omega_{\rm NL} \to 1$, it follows that 
\be
r_H=\sqrt{57}/6-1/2\,,\qquad
w_{\rm DE} \simeq -1.506\,.
\ee
Since $r_H$ is constant, the growth rates of the
Hubble parameter squared $H^2$ are the same as
those of $V$ and $S$.
Hence we have the super-inflationary solution 
$H \propto a^{\sqrt{57}/6-1/2}$ approaching 
a big-rip singularity. However the above study 
neglects the contribution of the $O(R^2)$ terms, so 
inclusion of those terms can modify the cosmological 
dynamics in the high-curvature regime.

\begin{figure}
\includegraphics[height=3.2in,width=3.3in]{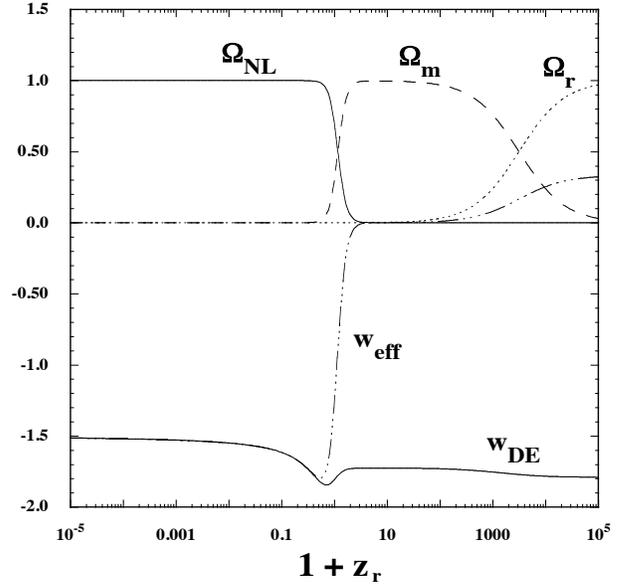}
\caption{\label{fig1}
Evolution of $\Omega_{\rm NL}$, $\Omega_m$, $\Omega_r$, 
$w_{\rm DE}$, and $w_{\rm eff}$ versus the redshift 
$z_r$ for the initial conditions $U=U'=0$, $V=V'=0$, 
$S=S'=0$, $\Omega_r=0.9992$, and $H/m=1.0 \times 10^{18}$ 
at $z_r=3.9 \times 10^6$. 
There is no cosmological constant in this simulation.
The present epoch ($z=0$, $H=H_0$) 
is identified as $\Omega_{\rm NL}=0.7$. 
In this case the mass $m$ corresponds to 
$m/H_0=1.5 \times 10^{-7}$.}
\end{figure}

In order to confirm the above analytic estimation,  
we numerically integrate
Eqs.~(\ref{auto1})-(\ref{auto5}) with the initial conditions 
$U=U'=0$, $V=V'=0$, and $S=S'=0$ in the deep radiation era.
In Fig.~\ref{fig1} we plot the evolution of $w_{\rm DE}$ 
and $w_{\rm eff}$ as well as the density parameters 
$\Omega_{\rm NL}$, $\Omega_m$, $\Omega_r$ 
versus the redshift $z_r=1/a-1$.
Clearly there is the sequence of radiation ($\Omega_r \simeq 1$, 
$w_{\rm eff} \simeq 1/3$), matter ($\Omega_m \simeq 1$, 
$w_{\rm eff}=0$), and dark energy 
($\Omega_{\rm NL} \simeq 1$, $w_{\rm eff}\simeq -1.5$) 
dominated epochs. 
We identify the present epoch ($z_r=0$) to be 
$\Omega_{\rm NL}=0.7$.
As we estimated analytically, the dark energy equation 
of state evolves as $w_{\rm DE} \simeq -1.791$ (radiation era), 
$w_{\rm DE} \simeq -1.725$ (matter era), and 
$w_{\rm DE} \simeq -1.506$ (accelerated era).

Notice that, even with the initial conditions $V=V'=0$, 
the growing-mode solution to Eq.~(\ref{auto2}) cannot be 
eliminated due to the presence of the term $-2r_H$.
Taking into account the decaying-mode solution to $V$ 
in the radiation era, 
the coefficient $c_2$ of Eq.~(\ref{Vso}) corresponding to $V=V'=0$ at 
$N=0$ (i.e., $a=a_i$) is $c_2=\sqrt{33}/132+1/4\simeq 0.29$.
Up to the radiation-matter equality ($a=a_{\rm eq}$),  
the field evolves as $V \simeq 0.29 (a/a_i)^{(\sqrt{33}-1)/2}$.
Since $V$ is proportional to $a^{(\sqrt{137}-3)/4}$ in the matter 
era, it follows that
\bee
V \simeq 0.29 
\left( \frac{ a_{\rm eq}}{a_i }\right)^{\frac{\sqrt{33}-1}{2}}
\left(\frac{a}{a_{\rm eq}}\right)^{\frac{\sqrt{137}-3}{4}}\,,
\label{Ves}
\eee
for $a_{\rm eq}<a<1$.

In the numerical simulation of Fig.~\ref{fig1} the initial condition 
is chosen to be $a_i=2.6 \times 10^{-7}$ with $a_{\rm eq}= 3.1 \times 10^{-4}$.
Since the cosmic acceleration starts when the last term in Eq.~(\ref{rH})
grows to the order of 1, we have
$m^2V_0/(8H_0^2) \approx 1$, where $V_0$ is the today's value of $V$.
Using the analytic estimation (\ref{Ves}), the mass $m$ is constrained to 
be $m \approx 10^{-7}H_0$.
In fact, this is close to the numerically derived value 
$m=1.5 \times 10^{-7}H_0$.

Thus, the mass $m$ is required to be much smaller 
than $H_0$ to avoid the early beginning of cosmic acceleration.
If the onset of the radiation era occurs at the redshift $z_r$ larger than 
$10^{15}$, i.e., $a_i \lesssim 10^{-15}$, the analytic estimation (\ref{Ves}) 
shows that the mass $m$ needs to satisfy the condition 
$m \lesssim 10^{-17}H_0$ to realize the successful cosmic
expansion history.

If we consider the evolution of the Universe earlier than the 
radiation era (e.g., inflation), the upper bound of $m$ 
should be even tighter. 
On the de Sitter background ($\dot{H}=0$) we have 
$r_H=0$, in which case the growth of $V$ can be 
avoided for the initial conditions $V=V'=0$. 
However, inflation in the early Universe has a small 
deviation from the exact de Sitter solution \cite{inflation} and hence the field 
$V$ can grow at some extent due to the non-vanishing 
values of $r_H$. For the theoretical consistency we need 
to include the $O(R^2)$ terms in such a high-energy regime, 
which is beyond the scope of our paper.

\subsection{$\Omega_{\Lambda} \neq 0$}
\begin{figure}
\includegraphics[height=3.2in,width=3.3in]{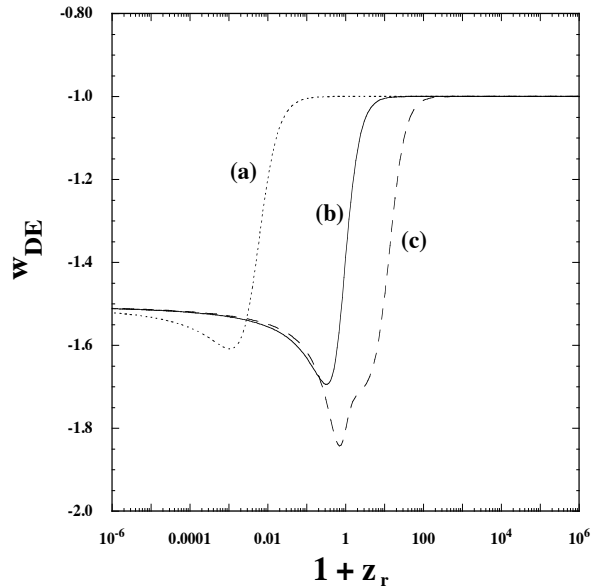}
\caption{\label{fig2}
Evolution of $w_{\rm DE}$ versus the redshift $z_r$ 
for the initial conditions 
(a) $\Omega_r=0.9965$, $\Omega_{\Lambda}=1.0 \times 10^{-20}$, 
$H/m=1.0 \times 10^{18}$ at $z_r=9.3 \times 10^5$,  
(b) $\Omega_r=0.999$, $\Omega_{\Lambda}=2.3 \times 10^{-23}$, 
$H/m=1.0 \times 10^{18}$ at $z_r=3.7 \times 10^6$,  
and (c) $\Omega_r=0.999$, $\Omega_{\Lambda}=1.0 \times 10^{-25}$, 
$H/m=1.0 \times 10^{18}$ at $z_r=4.1 \times 10^6$.
In all these cases the initial conditions of the fields are chosen 
to be $U=U'=0$, $V=V'=0$, and $S=S'=0$.
We identify the present epoch at 
$\Omega_{\rm NL}+\Omega_{\Lambda}=0.7$. }
\end{figure}
\begin{figure}
\includegraphics[height=3.2in,width=3.3in]{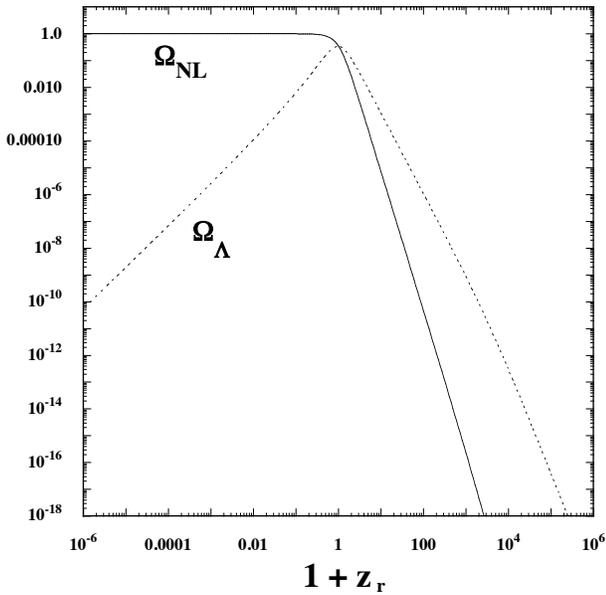}
\caption{\label{fig3}
Evolution of $\Omega_{\rm NL}$ and $\Omega_{\Lambda}$ 
versus the redshift $z_r$ for the initial conditions 
corresponding to the case (b) in Fig.~\ref{fig2}. }
\end{figure}

In the presence of the cosmological constant with the 
energy density $\rho_{\Lambda}$, 
the cosmological dynamics is subject to change relative to 
that studied in Sec.~\ref{nosec}.
During the radiation and matter eras we have
$3H^2 \gg |(m^2/4)(U+3V-4\dot{S}_0)|$ in Eq.~(\ref{be1}) 
and hence $3H^2 \simeq 8\pi G \rho$.
In order to avoid the appearance of $\rho_{\Lambda}$
in these epochs, we require the condition 
$\rho_{\Lambda} \lesssim 3H_0^2/(8\pi G)$.

The non-local mass term finally dominates over 
the cosmological constant because the equation of 
state of the former is smaller than that of the latter. 
If the condition $8 \pi G \rho_{\Lambda} \gg 
|(m^2/4)(U+3V-4\dot{S}_0)|$ is satisfied {\it today}, 
the non-local term comes out in the future.
The case (a) in Fig.~\ref{fig2} corresponds to such 
an example. In this case, the dark energy equation of 
state is close to $-1$ 
up to $z_r \sim -0.9$. It then approaches 
the asymptotic value $w_{\rm DE}=-1.506$.

For smaller values of $\rho_{\Lambda}$, the dominance 
of the non-local term occurs earlier. 
In the case (b) of Fig.~\ref{fig2} the energy densities of
the non-local term and the cosmological constant are
the same orders today ($\Omega_{\rm NL}=0.36$ 
and $\Omega_{\Lambda}=0.34$ at $z_r=0$).
In this case the dark energy equation of state starts 
to decrease only recently with the today's value 
$w_{\rm DE}=-1.39$.

In the case (c) the transition to the asymptotic regime 
$w_{\rm DE}=-1.506$ occurs even earlier
(around $z_r \sim 100$).
Observationally it is possible to distinguish between
the three different cases of Fig.~\ref{fig2}.
In the limit that $\rho_{\Lambda} \to 0$, the evolution of 
$w_{\rm DE}$ approaches the one shown 
in Fig.~\ref{fig1}.
For smaller $\rho_{\Lambda}$ the graviton mass $m$ tends 
to be larger because of the earlier dominance of 
the non-local term. For the cases (a), (b), and (c), 
the numerical values of the mass are 
$m=8.37 \times 10^{-9}H_0$, 
$m=1.22 \times 10^{-7}H_0$, and 
$m=1.46 \times 10^{-7}H_0$, respectively.

In Fig.~\ref{fig3} we plot the evolution of $\Omega_{\rm NL}$ 
and $\Omega_{\Lambda}$ for the initial conditions 
corresponding to the case (b) in Fig.~\ref{fig2}.
After $\Omega_{\rm NL}$ gets larger than $\Omega_{\Lambda}$ 
today, $\Omega_{\rm NL}$ approaches 1, 
while $\Omega_{\Lambda}$ starts to decrease 
toward 0. This behavior comes from the fact that, 
after the dominance of $\Omega_{\rm NL}$,
the terms on the left hand side of Eq.~(\ref{be1}) balance 
with each other, i.e., 
$3H^2+(m^2/4)(U+3V-4\dot{S}_0) \simeq 0$.
Then the cosmological constant appearing 
on the right hand side of Eq.~(\ref{be1}) effectively 
decouples from the dynamics of the system.
This is a kind of degravitation, by which the contribution of 
the matter component present in the energy density $\rho$ 
becomes negligible relative to that of the non-local term.

\section{Conclusions and discussions}
\label{consec}

In this paper we showed that the field equation of motion 
in the non-local massive gravity theory proposed 
by Jaccard {\it et al.} \cite{Maggiore}
follows from the covariant non-local Lagrangian (\ref{NLL}) with 
quadratic curvature terms. 
This is the generalization of the super-renormalizable 
massless theory with the ultraviolet modification factor 
$e^{{\mathrm H} (-\Box_M)}$.

Expanding the Lagrangian (\ref{NLL}) up to second order of 
the perturbations $h_{\mu \nu}$ on the Minkowski 
background, the propagator of the theory can be expressed
in terms of four operators which project out the spin-2, spin-1, 
and two spin-0 parts of a massive tensor field. 
The propagator (\ref{propgauge2b}) smoothly connects
to that of the massless theory in the limit $m \to 0$ and 
hence there is no vDVZ discontinuity. 
We also found that the theory described by (\ref{NLL}) 
is unitary at tree level, by coupling the propagator 
to external conserved stress-energy tensors and 
evaluating the residue of the amplitude at the pole ($k^2=m^2$).

In the presence of a conserved energy-momentum tensor 
$T_{\mu \nu}$, the non-local equation of motion following from 
the Lagrangian (\ref{NLL}) is given by Eq.~(\ref{EMT}).
In the low-energy regime much below the Planck scale 
the quadratic curvature terms can be negligible relative to 
other terms, so that the equation of motion reduces 
to (\ref{EMT2}) for ${\mathrm H}(-\Box_{M})=0$. 
We studied the cosmological dynamics based on 
the non-local equation (\ref{EMT2}) in detail on 
the flat FLRW background.

The tensor field $S_{\mu \nu}$, which satisfies the relation 
(\ref{Smure}), can be decomposed into the form (\ref{Smunu}). 
In order to respect the continuity equation 
$\nabla^{\mu} T_{\mu \nu}=0$ for matter, the transverse 
part of $S_{\mu \nu}$ needs to be extracted in the 
second term on the left hand side of Eq.~(\ref{EMT2}). 
Among the components of the vector $S_{\mu}$ 
in Eq.~(\ref{Smunu}), the three vector $S_i$ ($i=1,2,3$) 
vanishes because of the symmetry of the FLRW space-time.
In addition to the vector component $S_0$, we also have 
two scalar degrees of freedom $U=S^0_0+S^i_i$ 
and $V=S^0_0-S^i_i/3$.

Among these dynamical degrees of freedom, the scalar 
field $V$ exhibits instabilities for the cosmological 
background with $\dot{H} \neq 0$.
Even in the absence of a dark energy component,
a late-time accelerated expansion of the Universe 
can be realized by the growth of $V$. 
In order to avoid an early entry to the phase of
cosmic acceleration, the graviton mass $m$ is required 
to be very much smaller than the today's 
Hubble parameter $H_0$.
We showed that the equation of state of this 
``dark'' component evolves as 
$w_{\rm DE}=-1.791$ (radiation era), 
$w_{\rm DE}=-1.725$ (matter era), and 
$w_{\rm DE}=-1.506$ (accelerated era), 
see Fig.~\ref{fig1}.
 
While the above property of the non-local massive gravity 
is attractive, the evolution of $w_{\rm DE}$ smaller than  
$-1.5$ during the matter and accelerated epochs is 
in tension with the joint data analysis of SNIa, CMB, 
and BAO \cite{obser}. 
In the presence of the cosmological constant $\Lambda$ 
(or other dark energy components such as quintessence),
the dark energy equation of state can evolve with 
the value close to $w_{\rm DE}=-1$ in the deep matter era 
(see Fig.~\ref{fig2}). In such cases the model can be consistent 
with the observational data. 
In the asymptotic future the non-local term dominates over 
the cosmological constant, which can be regarded as a kind 
of degravitation of $\Lambda$.

Recently, Maggiore \cite{Maggiore13} studied the modified version 
of the non-local massive gravity in which the second term on the 
left hand side of Eq.~(\ref{EMT2}) is replaced by 
$m^2 (g_{\mu \nu} \Box^{-1}R)^{\rm T}$, where T denotes
the extraction of the transverse part. 
In this theory the $-2\dot{H}$ term on the right hand side of Eq.~(\ref{ddotv2}) 
disappears, in which case the growth of $V$ can be avoided 
for the initial conditions $V=\dot{V}=0$ (i.e., decoupled from 
the dynamics). Since the growth of the fields $U$ and 
$S_0$ is milder than that of the field $V$ studied in Sec.~\ref{nosec}, 
$w_{\rm DE}$ evolves from the value slightly smaller than $-1$ 
during the matter era to the value larger than $-1$ \cite{Maggiore13}.
It will be of interest to study whether such a theory can be consistently 
formulated in the framework of the covariant action related 
to the super-renormalizable massless theory.

While we showed that the theory described by the covariant 
Lagrangian (\ref{NLL}) is tree-level unitary on the 
Minkowski background, it remains to see what happens 
on the cosmological background. 
This requires detailed study for the expansion 
of the Lagrangian (\ref{NLL}) up to second order 
in cosmological perturbations about the FLRW background.
We leave such analysis for future work.

\section*{ACKNOWLEDGEMENTS}
L.~M. and S.~T. are grateful to Gianluca Calcagni for the invitations
to 1-st i-Link workshop on quantum gravity and cosmology at which this 
project was initiated.
S.~T. is supported by the Scientific Research Fund of the 
JSPS (No.~24540286) and financial support from Scientific Research 
on Innovative Areas (No.~21111006). 


\begin{thebibliography}{99}

\bibitem{Fierz} 
M.~Fierz, Helv.\ Phys.\ Acta \textbf{12}, 3 (1939);
M.~Fierz and W.~Pauli, Proc.\ Roy.\ Soc. \ Lond. A \textbf{173},
211 (1939).

\bibitem{DVZ} 
H.~van Dam and M.\ J.\ G.~Veltman, Nucl.\ Phys. \, B \textbf{22}, 
397 (1970); 
V.\ I.~Zakharov, JETP Lett. \ \textbf{12}, 312 (1970); 
Y.~Iwasaki, 
Phys.\ Rev.\ \textbf{D2}, 2255-2256 (1970).

\bibitem{Vain} 
A.~I.~Vainshtein, 
Phys.\ Lett.\ B \textbf{39}, 393 (1972).

\bibitem{BDghost} 
D.~G.~Boulware and S.~Deser, Phys.\ Rev.\ \textbf{D6},
3368 (1972).

\bibitem{dRGT} 
C.~de Rham, G.~Gabadadze and A.~J.~Tolley,
Phys.\ Rev.\ Lett.\  {\bf 106}, 231101 (2011)
[arXiv:1011.1232 [hep-th]].

\bibitem{Galileon} 
A.~Nicolis, R.~Rattazzi and E.~Trincherini,
Phys.\ Rev.\ D {\bf 79}, 064036 (2009)
[arXiv:0811.2197 [hep-th]];
C.~Deffayet, G.~Esposito-Farese and A.~Vikman,
Phys.\ Rev.\ D {\bf 79}, 084003 (2009)
[arXiv:0901.1314 [hep-th]].

\bibitem{ins} 
A.~E.~Gumrukcuoglu, C.~Lin and S.~Mukohyama,
JCAP {\bf 1111}, 030 (2011)
[arXiv:1109.3845 [hep-th]];
JCAP {\bf 1203}, 006 (2012)
[arXiv:1111.4107 [hep-th]];
K.~Koyama, G.~Niz and G.~Tasinato,
Phys.\ Rev.\ D {\bf 84}, 064033 (2011)
[arXiv:1104.2143 [hep-th]].

\bibitem{Antonio} 
A.~De Felice, A.~E.~Gumrukcuoglu and S.~Mukohyama,
Phys.\ Rev.\ Lett.\  {\bf 109}, 171101 (2012)
[arXiv:1206.2080 [hep-th]].

\bibitem{aca} 
S.~Deser and A.~Waldron,
Phys.\ Rev.\ Lett.\  {\bf 110}, 111101 (2013)
[arXiv:1212.5835 [hep-th]].

\bibitem{quasi1}
G.~D'Amico, G.~Gabadadze, L.~Hui and D.~Pirtskhalava,
Phys.\ Rev.\ D {\bf 87}, 064037 (2013)
[arXiv:1206.4253 [hep-th]].
 
\bibitem{quasi2}
Q.~-G.~Huang, Y.~-S.~Piao and S.~-Y.~Zhou,
Phys.\ Rev.\ D {\bf 86}, 124014 (2012)
[arXiv:1206.5678 [hep-th]].

\bibitem{quasi3}
A.~De Felice and S.~Mukohyama,
arXiv:1306.5502 [hep-th].

\bibitem{homoge}
G.~D'Amico {\it et al}.,
Phys.\ Rev.\ D {\bf 84}, 124046 (2011)
[arXiv:1108.5231 [hep-th]].

\bibitem{Gum}
A.~E.~Gumrukcuoglu, C.~Lin and S.~Mukohyama,
Phys.\ Lett.\ B {\bf 717}, 295 (2012)
[arXiv:1206.2723 [hep-th]].

\bibitem{Antonio2}
A.~De Felice, A.~E.~Gumrukcuoglu, C.~Lin and S.~Mukohyama,
JCAP {\bf 1305}, 035 (2013)
[arXiv:1303.4154 [hep-th]].

\bibitem{Maggiore}	
M.~Jaccard, M.~Maggiore and E.~Mitsou,
Phys.\ Rev.\ D {\bf 88}, 044033 (2013)
[arXiv:1305.3034 [hep-th]].

\bibitem{Arkani} 
N.~Arkani-Hamed, S.~Dimopoulos, G.~Dvali and G.~Gabadadze,
hep-th/0209227.

\bibitem{Dvali} 
G.~Dvali, S.~Hofmann and J.~Khoury,
Phys.\ Rev.\ D {\bf 76}, 084006 (2007)
[hep-th/0703027].

\bibitem{M1} 
L.~Modesto,
Phys.\ Rev.\ D {\bf 86}, 044005 (2012)
[arXiv:1107.2403 [hep-th]]; 
L.~Modesto,
arXiv:1305.6741 [hep-th].

\bibitem{Krasnikov} 
N.~V.~Krasnikov,
Theor.\ Math.\ Phys.\  {\bf 73}, 1184 (1987)
[Teor.\ Mat.\ Fiz.\  {\bf 73}, 235 (1987)].

\bibitem{Tomboulis} 
E.~T.~Tomboulis,
hep-th/9702146.

\bibitem{BM} 
T.~Biswas, E.~Gerwick, T.~Koivisto and A.~Mazumdar,
Phys.\ Rev.\ Lett.\  {\bf 108}, 031101 (2012)
[arXiv:1110.5249 [gr-qc]]; 

\bibitem{M2}	
L.~Modesto,
arXiv:1202.3151 [hep-th]; 
L.~Modesto,
arXiv:1202.0008 [hep-th].

\bibitem{M3} 
S.~Alexander, A.~Marciano and L.~Modesto,
Phys.\ Rev.\ D {\bf 85}, 124030 (2012)
[arXiv:1202.1824 [hep-th]].
  
\bibitem{M4} 
F.~Briscese, A.~Marciano, L.~Modesto and E.~N.~Saridakis,
Phys.\ Rev.\ D {\bf 87}, 083507 (2013)
[arXiv:1212.3611 [hep-th]].
  
\bibitem{Deser} 
S.~Deser and R.~P.~Woodard,
Phys.\ Rev.\ Lett.\  {\bf 99}, 111301 (2007)
[arXiv:0706.2151 [astro-ph]].

\bibitem{Jhingan} 
S.~Jhingan {\it et al.},
Phys.\ Lett.\ B {\bf 663}, 424 (2008)
[arXiv:0803.2613 [hep-th]].

\bibitem{Koivisto} 
T.~Koivisto,
Phys.\ Rev.\ D {\bf 77}, 123513 (2008)
[arXiv:0803.3399 [gr-qc]];
Phys.\ Rev.\ D {\bf 78}, 123505 (2008)
[arXiv:0807.3778 [gr-qc]]. 

\bibitem{Woodard} 
C.~Deffayet and R.~P.~Woodard,
JCAP {\bf 0908}, 023 (2009)
[arXiv:0904.0961 [gr-qc]].

\bibitem{Zhang} 
Y.~-l.~Zhang and M.~Sasaki,
Int.\ J.\ Mod.\ Phys.\ D {\bf 21}, 1250006 (2012)
[arXiv:1108.2112 [gr-qc]].

\bibitem{Elizalde} 
E.~Elizalde, E.~O.~Pozdeeva and S.~Y.~.Vernov,
Phys.\ Rev.\ D {\bf 85}, 044002 (2012)
[arXiv:1110.5806 [astro-ph.CO]];
E.~Elizalde, E.~O.~Pozdeeva, S.~Y.~Vernov and Y.~-l.~Zhang,
JCAP {\bf 1307}, 034 (2013)
[arXiv:1302.4330 [hep-th]].

\bibitem{Park} 
S.~Park and S.~Dodelson,
Phys.\ Rev.\ D {\bf 87}, 024003 (2013)
[arXiv:1209.0836 [astro-ph.CO]].
  
\bibitem{HigherDG}
A.~Accioly, A.~Azeredo and H.~Mukai,
J.\ Math.\ Phys.\  {\bf 43}, 473 (2002).
  
\bibitem{Stelle} 
K.~S.~Stelle,
Phys.\ Rev.\ D {\bf 16}, 953 (1977).

\bibitem{VanNieuwenhuizen:1973fi} 
P.~Van Nieuwenhuizen,
Nucl.\ Phys.\ B {\bf 60}, 478 (1973).
  
\bibitem{Accioly:2011nf} 
A.~Accioly, J.~Helayel-Neto, E.~Scatena, J.~Morais, R.~Turcati and B.~Pereira-Dias,
Class.\ Quant.\ Grav.\  {\bf 28}, 225008 (2011)
[arXiv:1108.0889 [hep-th]].
 
\bibitem{Veltman:1975vx} 
M.~J.~G.~Veltman,
Conf.\ Proc.\ C {\bf 7507281}, 265 (1975).

\bibitem{barvy} 	
A.~O.~Barvinsky,
Phys.\ Lett.\ B {\bf 572}, 109 (2003)
[hep-th/0304229].
  
\bibitem{barvy2}
A.~O.~Barvinsky, A.~Y.~Kamenshchik, A.~Rathke and C.~Kiefer,
Phys.\ Rev.\ D {\bf 67}, 023513 (2003)
[hep-th/0206188].

\bibitem{barvy3}
A.~O.~Barvinsky and Y.~.V.~Gusev,
Phys.\ Part.\ Nucl.\  {\bf 44}, 213 (2013)
[arXiv:1209.3062 [hep-th]].

\bibitem{barvy4}
A.~O.~Barvinsky,
Phys.\ Lett.\ B {\bf 710}, 12 (2012)
[arXiv:1107.1463 [hep-th]].
  
\bibitem{smolin}
N.~Afshordi,
arXiv:0807.2639 [astro-ph];
L.~Smolin,
Phys.\ Rev.\ D {\bf 80}, 084003 (2009)
[arXiv:0904.4841 [hep-th]].
  
\bibitem{Porrati} 
M.~Porrati,
Phys.\ Lett.\ B {\bf 534}, 209 (2002)
[hep-th/0203014].  

\bibitem{inflation} 
P.~A.~R.~Ade {\it et al.}  [Planck Collaboration],
arXiv:1303.5076 [astro-ph.CO];
S.~Tsujikawa, J.~Ohashi, S.~Kuroyanagi and A.~De Felice,
Phys.\  Rev.\ D {\bf 88}, 023529 (2013)
[arXiv:1305.3044 [astro-ph.CO]].
    
\bibitem{obser}
S.~Nesseris, A.~De Felice and S.~Tsujikawa,
Phys.\ Rev.\ D {\bf 82}, 124054 (2010)
[arXiv:1010.0407 [astro-ph.CO]];
A.~De Felice and S.~Tsujikawa,
JCAP {\bf 1203}, 025 (2012)
[arXiv:1112.1774 [astro-ph.CO]];
G.~Hinshaw {\it et al.}  [WMAP Collaboration],
arXiv:1212.5226 [astro-ph.CO].
  
\bibitem{Maggiore13} 
M.~Maggiore,
arXiv:1307.3898 [hep-th].


\end{thebibliography}
\end{document}